\documentstyle[aps,prl,twocolumn,floats,epsf]{revtex}
\begin{document}
\draft
\title{Diagonalization of system plus environment Hamiltonians}
\author{Stefan K.~Kehrein$^1$ and Andreas Mielke$^2$}
\address{Theoretische Physik III --- Elektronische Korrelationen und
Magnetismus$^1$, Institut f\"ur Physik,\\
Universit\"at~Augsburg, 86135~Augsburg, Germany \\
Institut f\"ur Theoretische Physik$^2$,
Ruprecht--Karls--Universit\"at, Philosophenweg~19,
69120~Heidelberg, Germany \\
{\rm(January 14, 1997)}
}

\address{~
\parbox{15cm}{\rm
\medskip
A new approach to dissipative quantum systems modelled by
a system plus environment Hamiltonian is presented. Using
a continuous sequence of
infinitesimal unitary transformations the small quantum
system is decoupled from its thermodynamically large
environment. Dissipation enters through the observation
that system observables generically ``decay"ï
completely into a different structure when the Hamiltonian
is transformed into diagonal form. The method is particularly
suited for studying low--temperature properties. This
is demonstrated explicitly for the super--Ohmic spin--boson model.
\\~\\
PACS numbers: 05.30.-d, 66.30.Dn, 05.40.+j
}}

\maketitle

\narrowtext
Dissipative behaviour in quantum physics can be modelled
in a system plus bath framework\cite{CL}. The quantum system that
one is interested in is described by a Hamiltonian~$H_S$ operating
on the system Hilbert space~${\cal H}_S$, the thermodynamically large
environment by some Hamiltonian~$H_B$ operating on~${\cal H}_B$.
The full Hamiltonian
$H: {\cal H}_S\otimes{\cal H}_B\rightarrow {\cal H}_S\otimes{\cal H}_B$
of system plus bath is obtained by
coupling ${\cal H}_S$ and ${\cal H}_B$ with some interaction~$H_{SB}$
\begin{equation}
\label{initial_H}
H=H_S\otimes\openone_B + \openone_S\otimes H_B + H_{SB}.
\end{equation}
Most theoretical work starts off by tracing out the bath degrees
of freedom and then using suitable approximation schemes for the
time evolution of the reduced density matrix of the small quantum
system (for a review see e.g. Ref.\cite{Weiss}).
In this letter we present an alternative approach that
aims at decoupling system and bath with a unitary transformation~$U$
\begin{equation}
\label{final_H}
U\,H\,U^\dag=\tilde H_S \otimes\openone_B + \openone_S\otimes \tilde H_B .
\end{equation}
Here $\tilde H_S$ and $\tilde H_B$ are modified system and
bath Hamiltonians. By carrying out this programme in the manner
described below this approach is particularly
suited for studying low--temperature properties of
dissipative quantums systems, thereby being complementary to
most other approximation schemes. As a specific
example we demonstrate these ideas for the spin--boson model
\begin{equation}
\label{spinboson}
H=-\frac{\Delta}{2}\sigma_x+\sum_k \omega_k b^\dag_k b_k
+\frac{1}{2}\sigma_z\sum_k \lambda_k (b^\dag_k+b_k)
\end{equation}
describing a two--level system coupled to a bath modelled
by harmonic oscillators. The standard approach to this
problem is the ``Non--Interacting Blip Approximation'' (NIBA)
for the effective action obtained after integrating out the
bath degrees of freedom \cite{NIBA}.
In our approach we find low--temperature
equilibrium correlation functions of the tunneling particle
that combine NIBA--results at intermediate time scales
with the correct long--time behaviour where the simple NIBA
fails \cite{Weiss}.
The universal Wilson ratio for a super--Ohmic bath put forward
in Ref.\cite{SW} is also obtained.

Extension of our scheme to other dissipative
quantum systems is straightforward under the basic
assumption that $H_S$ has a non--degenerate ground state
separated by a finite gap from its excited states.
Some technical details of our method can be found in
Ref.\cite{KM}.

Two obvious questions arise with respect to the programme
(\ref{initial_H})--(\ref{final_H}). Where is dissipation
in Eq.~(\ref{final_H}) as exchange of energy between system
and bath is no longer possible? And how can one find a unitary
transformation~$U$ that fulfills the required task?

We first discuss the second point.
The decoupling of system and bath is achieved by a method
of infinitesimal unitary transformations
(``flow equations'') introduced by
Wegner in Ref.\cite{W}. A suitable antihermitean generator
$\eta(\ell)=-\eta(\ell)^\dag$ is chosen and the initial--value
problem
\begin{equation}
\label{unitarytrf}
\frac{dH(\ell)}{d\ell}=[\eta(\ell),H(\ell)] ,\quad H(\ell =0)=H
\end{equation}
solved. The parameter $\ell$ has dimension (Energy)$^{-2}$.
Eq.~(\ref{unitarytrf}) generates a one--parameter family of
unitarily equivalent Hamiltonians $H(\ell)$.
In the limit~$\ell\rightarrow\infty$ one attempts to
obtain a Hamiltonian $H(\ell=\infty)$ of the simple structure
(\ref{final_H}). The choice of $\eta(\ell)$ is inspired by
renormalization theory: Initially for small~$\ell$
matrix elements corresponding to large energy
differences between system and bath are decoupled, for
large~$\ell$ one deals with the nearly resonant modes. The
fundamental problem of (\ref{unitarytrf}) is that higher and
higher order interactions are successively generated. So a
further condition for~$\eta(\ell)$ is that the number of
additional terms should be small. These conditions essentially
fix a unique structure of~$\eta(\ell)$
\begin{eqnarray}
\eta&=&\imath\sigma_y\sum_k \eta^{(y)}_k (b_k+b^\dag_k)
+\sigma_z \sum_k\eta^{(z)}_k (b_k-b^\dag_k) \nonumber \\
&&+\sum_{k,q} \eta_{k,q} :(b_k+b^\dag_k)(b_q-b^\dag_q):
\end{eqnarray}
with
\begin{eqnarray}
\eta^{(y)}_k&=&-\frac{1}{2}\lambda_k\Delta
\frac{\omega_k-\Delta}{\omega_k+\Delta} ,\quad
\eta^{(z)}_k=\frac{\omega_k}{\Delta}\eta^{(y)}_k ,
\label{eta_para} \\
\eta_{k,q}&=&\frac{\lambda_k\lambda_q\Delta\omega_q}{
2(\omega_k^2-\omega_q^2)} \tanh\frac{\beta\Delta}{2}
\left(\frac{\omega_k-\Delta}{\omega_k+\Delta}
+\frac{\omega_q-\Delta}{\omega_q+\Delta}\right) .\nonumber
\end{eqnarray}
All parameters in (\ref{eta_para}) depend on $\ell$ and
normal--ordering with respect to the noninteracting
ground state has been introduced. For
the construction of generators~$\eta$ in a general setting
see Refs.~\cite{KM,W}.
The only new interaction terms generated in the first application of
(\ref{unitarytrf}) is $:\sigma_x(b_k\pm b^\dag_k)
(b_{k'}\pm b^\dag_{k'}):$.

Due to the flow the spectral function
$
J(\omega)=\sum_k \lambda_k^2\,\delta(\omega-\omega_k)
$
describing the coupling of system and bath becomes $\ell$--dependent
too, $J(\omega,\ell)=\sum_k \lambda_k(\ell)^2\,\delta(\omega-\omega_k)$,
$J(\omega,\ell=0)=J(\omega)$. We end up with the following set
of differential equations for the couplings in the
Hamiltonian by comparing the lhs and rhs of~(\ref{unitarytrf})
\begin{eqnarray}
\label{flowDelta}
\frac{d\Delta}{d\ell}&=&-\Delta\int d\omega\,J(\omega,\ell)
\frac{\omega-\Delta}{\omega+\Delta}
\coth\frac{\beta\omega}{2} , \\
\label{flowJ}
\frac{\partial J(\omega,\ell)}{\partial\ell}&=&-2(\omega-\Delta)^2
J(\omega,\ell)
+2\Delta\tanh\frac{\beta\Delta}{2} J(\omega,\ell) \nonumber \\
&&\times
\int d\omega'\,\frac{\omega'J(\omega',\ell)}{\omega^2-\omega'^2}
\left(\frac{\omega-\Delta}{\omega+\Delta}
+\frac{\omega'-\Delta}{\omega'+\Delta}\right)
\end{eqnarray}
and a differential equation for an
uninteresting additive term in (\ref{spinboson}). The differential
equation for the higher
normal--ordered interaction term $:\sigma_x(b_k\pm b^\dag_k)
(b_{k'}\pm b^\dag_{k'}):$ is subsequently
negelected. This approximation can be systematically improved by
taking such higher--order interactions that are also of higher
order in the small parameters~$\lambda_k$ into account one after the
other in the hierarchy of differential equations. However,
already the above approximation
leads to a very satisfactory description as will be shown.

As required large energy
differences are first decoupled in (\ref{flowJ}) and
small energy differences later. For $\ell\rightarrow\infty$
the coupling $J(\omega,\ell)$ vanishes for all $\omega$, in
general exponentially and for $\omega=\Delta(\ell=\infty)$
algebraically. Within the above approximations one ends up
with a system Hamiltonian $\tilde H_S=-1/2\,\Delta_r\,\sigma_x$
that is decoupled from the environment. Here
$\Delta_r=\Delta(\ell=\infty)$ is the renormalized
tunneling matrix element. By either numerical solution of
the differential equations or analytical approximations
one finds
$
\Delta_r=c\,\Delta_0
\exp\big(-(\omega_c/K)^{s-1}/(2s-2)\big)
$
for a super--Ohmic bath $J(\omega)=K^{1-s}\omega^s
\Theta(\omega_c-\omega)$ with $s>1$. Here $\Delta_0=\Delta(\ell=0)$,
$c$ is a constant of order~1, $\omega_c$ some high--energy
cutoff and $K$ the coupling constant. $\Delta_r$
defines the low--energy scale of the problem in agreement
with other methods (see e.g. Ref.\cite{NIBA}).

For small $\ell\ll \Delta_r^{-2}$ the above procedure
is equivalent to Anderson's ``poor man's'' scaling
\cite{A} with a smooth cutoff, except that the high--energy
states are not removed by integrating them out but by decoupling
them using unitary transformations (see Fig.~1).
By explicitly finding this
unitary transformation no information about the removed states
is lost but contained in the unitary transformation
itself\cite{global}. Scaling approaches
have to stop when the effective band edge becomes of order
the low--energy scale of the problem due to divergencies
in the renormalization group equations. The flow equations
can be integrated further since with our choice of
$f(\omega,\ell)$ decoupling is with respect to
energy differences and does not only take place at
the effective band edge.

\begin{figure}[t]
\begin{center}
\leavevmode
\epsfxsize=8.5cm
\epsfysize=5.7cm
\epsfbox{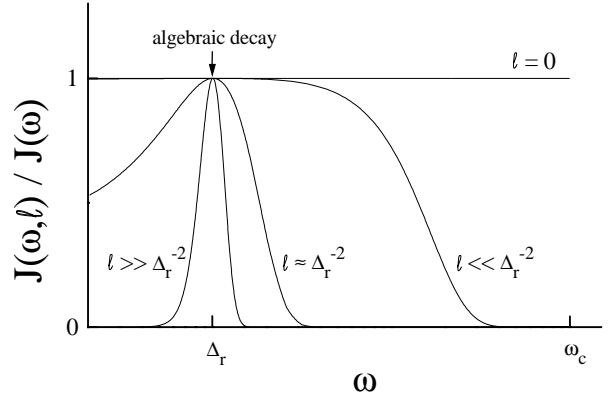}
\caption{Schematic behaviour of the
effective spectral function $J(\omega,\ell)$ for
various regimes of the flow equations.}
\end{center}
\end{figure}

In the region $\ell\gtrsim\Delta_r^{-2}$ that remains
unreachable in the ``poor man's'' approach new features
appear that are hidden for smaller~$\ell$. First of
all the flow of the parameters in the Hamiltonian becomes
negligible and has the universal algebraic
behaviour~$\ell^{-1/2}$. In contrast the transformation of
the system observables turns out to be essential in this regime.
In the crossover region
$\ell\approx\Delta_r^{-2}$ both flow of parameters
and system observables are important.

As a specific example for this scenario we discuss the
symmetrized equilibrium correlation
function describing the tunneling particle
$
C(t)=\frac{1}{2}\,<\{\sigma_z(t),\sigma_z(0)\}> .
$
In order to use the trivial time evolution with respect
to the Hamiltonian~$H(\ell=\infty)$ we have to transform the
observable $\sigma_z$ as well
\begin{equation}
\frac{d\sigma_z(\ell)}{d\ell}=[\eta(\ell),\sigma_z(\ell)],\quad
\sigma_z(\ell=0)=\sigma_z .
\end{equation}
These differential equations cannot be solved in closed form
and we have to make an ansatz for the transformation of
$\sigma_z(\ell)$
\begin{equation}
\label{flow_sigmaz}
\sigma_z(\ell)=h(l)\sigma_z+\sigma_x\sum_k\chi_k(\ell)
(b_k+b^\dag_k)
\end{equation}
where higher normal--ordered terms are neglected. One obtains
the following differential equations
\begin{eqnarray}
\label{flowh}
\frac{dh}{d\ell}&=&-\Delta\sum_k \lambda_k \chi_k
\frac{\omega_k-\Delta}{\omega_k+\Delta}
\coth\frac{\beta\omega_k}{2} \\
\label{flowchi}
\frac{d\chi_k}{d\ell}&=&\Delta\,h\,\lambda_k
\frac{\omega_k-\Delta}{\omega_k+\Delta}
+\Delta\lambda_k\tanh\frac{\beta\Delta}{2} \\
&&\times\sum_q \frac{\chi_q\lambda_q\omega_q}{
\omega_k^2-\omega_q^2} \left(
\frac{\omega_k-\Delta}{\omega_k+\Delta}
+\frac{\omega_q-\Delta}{\omega_q+\Delta}\right) .
\nonumber
\end{eqnarray}
One can prove $h(\ell=\infty)=0$ if $\Delta_r$ lies
in the support of $J(\omega)$. Therefore $\sigma_z(\ell)$ in
(\ref{flow_sigmaz}) decays completely under the sequence of
unitary transformations, which is essential
to see dissipative behaviour with a Hamiltonian like
(\ref{final_H}). In general one can show for
a system observable $O \otimes\openone_B$ that does not
commute with the algebra spanned by $[H_S\otimes\openone_B,H_{SB}]$:
If some excitation energy from the ground state
of $\tilde H_S$ lies in the support
of the spectral function then the observable decays completely
when transforming from (\ref{initial_H}) to (\ref{final_H})
in the sense that no such term of structure
$O \otimes\openone_B$ survives \cite{KM}. This result also holds
in the zero--temperature limit.

The one--sided Fourier transform $C(\omega)$ of the correlation
function $C(t)=\int_0^\infty d\omega\,C(\omega)\,\cos(\omega t)$
can be expressed as
$
C(\omega)=\sum_k \chi_k^2(\infty) \coth(\beta\omega_k/2)\,
\delta(\omega-\omega_k)
$
and the $\chi_k(\infty)$ have to be found numerically. This is simplified
by the following observation. For $\ell\rightarrow\infty$ the
remaining spectral function $J(\omega,\ell)$ is strongly
peaked around $\Delta_r$ (Fig.~1). In this limit the
exactly solved dissipative harmonic oscillator\cite{HR}
\begin{equation}
H=\Delta b^\dag b+\sum_k \omega_k b^\dag_k b_k
+(b+b^\dag)\sum_k \lambda_k (b_k+b^\dag_k)
\end{equation}
becomes equivalent to a two--state system as
the mean occupation number $<b^\dag\,b>$
of the dissipative harmonic oscillator at zero temperature
goes to zero with the width of the spectral function. The
corresponding set of differential equations
(\ref{flowDelta},\ref{flowJ},\ref{flowh},\ref{flowchi}) for the
dissipative harmonic oscillator can be solved in closed form
\cite{KM}. This solution is exact as no higher--order
terms in the Hamiltonian or the transformation of observables
appear. Formally this can be seen by introducing functions
$S_0(z,\ell)=\sum_k \omega_k\lambda_k^2/(z-\omega_k^2)$,
$S_1(z,\ell)=\sum_k \sqrt{\omega_k\Delta}\chi_k\lambda_k/(z-\omega_k^2)$,
$S_2(z,\ell)=\sum_k \chi_k^2/(z-\omega_k^2)$ and by showing
\begin{eqnarray}
S_2(z,\infty)&=&S_2(z,\ell_0)
-\frac{(h(\ell_0)+S_1(z,\ell_0))^2}{
\Delta(\ell_0)^2-z+\Delta(\ell_0) S_0(z,\ell_0)} \nonumber\\
&&\quad +O(\ell_0^{-1}) .
\label{approx_cons}
\end{eqnarray}
For the dissipative harmonic oscillator (\ref{approx_cons}) holds
exactly without the term $O(\ell_0^{-1})$.
Numerically the correlation
function $C(\omega)=-2\omega/\pi\,\Im\,S_2(\omega^2-i0_+,\infty)$
has been obtained by integrating the flow equations up to
some $\ell_0=(2\lambda\Delta_r)^{-2}$, $\lambda\ll 1$,
thereby obtaining $S_2(z,\ell_0)$ and then adding the second
term from (\ref{approx_cons}) that describes a dissipative
harmonic oscillator with the spectral function $J(\omega,\ell_0)$.
Therefore the resolution of the peak in $C(\omega)$ for
$\omega=\Delta_r$ is not limited by~$\ell_0^{-1/2}$.
By using the analogy with the dissipative harmonic
oscillator we are restricted to low temperatures
$T\ll\Delta_r$.

Some zero--temperature correlation functions obtained in
this manner are shown in Fig.~2. The final results vary
very little with~$\lambda$ as long as $\lambda\lesssim 0.5$.
This gives a posteriori justification of our approximations.
For $\ell\ll\Delta_r^{-2}$ the neglected
terms are irrelevant in the usual scaling sense, for
$\ell\gg\Delta_r^{-2}$ our equations are closed
due to the analogy with the dissipative harmonic oscillator.
As the final results for super--Ohmic baths
do not depend on where these two parts
are matched in the crossover region, it is reasonable to argue that
the approximations are also good for
$\ell\approx\Delta_r^{-2}$ \cite{ohmic}.

\begin{figure}[t]
\begin{center}
\leavevmode
\epsfxsize=8.5cm
\epsfysize=5.7cm
\epsfbox{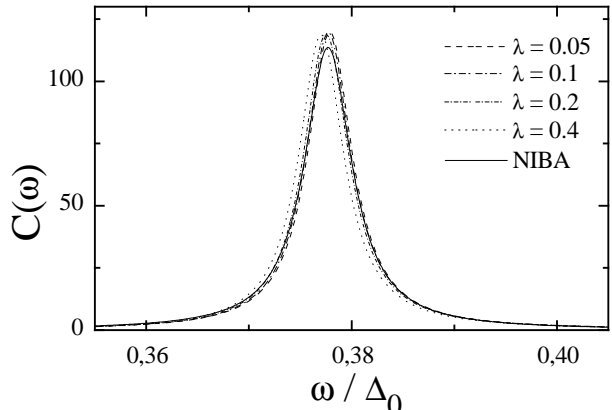}
\caption{Exemplary correlation function for a super--Ohmic bath
($s=2$, $K=40$, $\omega_c=80$). The different curves are obtained by
using Eq.~(\protect\ref{approx_cons}) for various values of
$\ell_0=(2\lambda\Delta_r)^{-2}$. The scale is set by
$\Delta_0=1$. For comparison the NIBA--curve
is also shown. The maximum of the NIBA--curve is slightly shifted
with respect to the flow equations and has been suitably rescaled
in order to identify the peaks.}
\end{center}
\end{figure}

The correlation functions obtained for such parameters are in
good agreement with the NIBA for intermediate time scales
(Fig.~2). That is for $\omega\approx\Delta_r$ the curves are
well--described by a Lorentzian with a peak at $\omega=\Delta_r$
and half width $\pi/2\,J(\Delta_r)$ \cite{analytic}.
For long times the NIBA fails as
it predicts an exponential decay at $T=0$ \cite{NIBA},
whereas the long--time behaviour is known to be
determined by the low--frequency behaviour of the spectral function.
This is made explicit in the Shiba--relation \cite{Shiba} generalized to
super--Ohmic baths in Ref.\cite{SW} (note our normalization
$\int_0^\infty C(\omega) d\omega=1$)
\begin{equation}
\lim_{\omega\rightarrow 0} \frac{C(\omega)}{(2\chi_0)^2 J(\omega)}
=1
\end{equation}
with the static susceptibility $\chi_0$. $\chi_0$
can be extracted with a Kramer's--Kronig
relation and a fluctuation-dissipation theorem
$\chi_0=1/2\int_0^\infty C(\omega)/\omega \,d\omega$. The numerical
solution of the flow equations is in excellent agreement with the
Shiba--relation (Table~1). One observes only small deviations
that disappear in the limit $\omega_c\gg\Delta_0$,
$\Delta_r={\rm const.}$ (or $\omega_c/K={\rm const.}$).

\begin{table}[t]
\squeezetable
\caption{Representative results from the numerical solution
of the flow equations
for super--Ohmic baths $J(\omega)=K^{1-s}\omega^s
\Theta(\omega_c-\omega)$. The scale is set by
$\Delta_0=1$. $R_s^{(\rm theor)}$ is
the Wilson ratio from (\protect\ref{Wilson}).
Numerical errors for the Shiba-- and the Wilson ratio
are estimated as~2\%.}
\begin{tabular}{cccccccc}
$s$ & $\omega_c$ & $K$ & $\chi_0$ &
$\left[\frac{C(\omega)}{J(\omega)}\right]_{\omega\rightarrow 0}$ &
$\left[\frac{C(\omega)}{(2\chi_0)^2 J(\omega)}\right]_{\omega\rightarrow 0}$ &
$R_s$ & $R_s^{(\rm theor)}$ \\ \hline
2 & 80 & 40 & 1.31 & 7.12 & 1.04 & 0.370 & 0.361 \\
2 & 160& 80 & 1.34 & 7.24 & 1.01 & 0.182 & 0.180 \\
2 & 320&160 & 1.35 & 7.31 & 1.00 & 0.090 & 0.090 \\
2 & 80 & 20 & 3.40 & 47.4 & 1.03 & 0.734 & 0.721 \\
2 & 160& 40 & 3.54 & 50.6 & 1.01 & 0.363 & 0.361 \\
2 & 320& 80 & 3.62 & 52.5 & 1.00 & 0.180 & 0.180 \\
2 & 80 & 10 & 22.6 & $2.05\times 10^3$ & 1.01 & 1.43  & 1.44 \\
2 & 160& 20 & 24.7 & $2.45\times 10^3$ & 1.00 & 0.718 & 0.721 \\
2 & 320& 40 & 25.9 & $2.69\times 10^3$ & 1.00 & 0.359 & 0.361 \\
3 & 20 & 10 & 1.26 & 6.68 & 1.05 & 0.786 & 0.779 \\
3 & 40 & 20 & 1.31 & 6.93 & 1.01 & 0.194 & 0.195 \\
3 & 60 & 30 & 1.33 & 7.05 & 1.00 & 0.0858& 0.0866 \\
3 & 10 & 3.33 & 3.52 & 52.4 & 1.06 & 7.39 & 7.03 \\
3 & 20 & 6.67 & 3.71 & 55.8 & 1.01 & 1.78 & 1.75 \\
3 & 30 & 10 & 3.98 & 63.8 & 1.01 & 0.783 & 0.779
\end{tabular}
\end{table}

Another interesting low--energy property is the impurity contribution
to the specific heat~$c_i(T)$. This quantity is trivial to obtain
from a Hamiltonian like (\ref{final_H}). But it follows from the difference
of two extensive quantities and it becomes necessary to discuss the flow
equations for the bath energies too. That could be ignored before
since the couplings~$\lambda_k$
scale with $1/\sqrt{N}$ where $N$ is the number of bath modes. One
finds
\begin{equation}
\frac{d\omega_k}{d\ell}=\Delta\,\lambda_k^2\,
\frac{\omega_k-\Delta}{\omega_k+\Delta}
\end{equation}
and the impurity contribution to the internal energy for
$T\ll \Delta_r$ is simply
\begin{equation}
E_i=\sum_k \frac{\omega_k(\infty)}{e^{\beta\omega_k(\infty)}-1}
-\sum_k \frac{\omega_k(0)}{e^{\beta\omega_k(0)}-1} .
\end{equation}
For super--Ohmic baths $J(\omega)\propto\omega^s$ for small
$\omega$, $s>1$, one derives the following expression for the
impurity contribution to the specific heat $c_i=\frac{dE_i}{dT}$
\begin{equation}
c_i=s\,T^s\,\Gamma(s+2)\,\zeta(s+1)\int_0^\infty d\ell\,
\Delta(\ell)\,g(\ell)
\end{equation}
with $g(\ell)$ defined from $J(\omega,\ell)=g(\ell)\,\omega^s
+O(\omega^{s+1})$. In particular the impurity contribution
to the specific heat is not Schottky--like but scales as $T^s$.
A sensitive test is provided by the Wilson ratio
$
R_s=\lim_{T\rightarrow 0} c_i(T)/(\chi_0\,T^s)
$
also generalized to super--Ohmic baths in Ref.\cite{SW}
\begin{equation}
\label{Wilson}
R_s=s\Gamma(s+2)\zeta(s+1)\lim_{x\rightarrow 0} \frac{J(x)}{x^s} .
\end{equation}
Wilson ratios obtained from the numerical solution of the
flow equations can be found in Table~1. Agreement
with (\ref{Wilson}) is excellent in the limit
$\omega_c\gg\Delta_0$, $\Delta_r={\rm const.}$

Summing up, we have applied a new approximation method based
on infinitesimal unitary transformations \cite{W} to the
spin--boson model. The method is an extension of
``poor man's'' scaling \cite{A} as it allows to
decouple modes below the low--energy scale of the model.
Instead of renormalization group equations with respect to
the effective band width we have differential equations with
respect to energy differences parametrized by~$\ell$. An essential
new feature as compared to the scaling approach is the
transformation of the observables once the decoupling
has reached the low--energy scale. For dissipative quantum systems
our method resulted in a complete decoupling of the quantum
system from its environment. Thereby we
have successfully matched formal solutions \cite{SW}
yielding the Shiba--relation and universal Wilson ratios
with the well--established NIBA at intermediate
energies in one consistent scheme.

%
%

%
%


\begin{references}
\bibitem{global} In principle it is
possible to reconstruct the full unitary transformation~$U$
in (\ref{final_H}) by evaluating $\ell$--ordered
products of $\eta(\ell)$, but the differential approach
is more appropriate.
\bibitem{ohmic}
For an Ohmic bath $J(\omega)\propto\omega$
with intermediate and strong coupling
the final result for $C(\omega)$ depends on~$\lambda$. This is due
to our choice of~$\eta_{k,q}$ that generates an IR--problem for
Ohmic baths \protect\cite{KM}. Hence we restrict ourselves to
super--Ohmic baths.
\bibitem{analytic}
This can also be shown analytically by an approximate solution of
the set of differential equations, see Ref.\protect\cite{KM}.
\end{references}
\end{document}